\documentclass{emulateapj}

\usepackage{verbatim}

\slugcomment{Last revision \today}

\newcommand{\gammat}{$\gamma_T$}
\newcommand{\gammacross}{$\gamma_\times$}
\newcommand{\deltasig}{$\Delta \Sigma$}

\newcommand{\deltarho}{$\Delta \rho$}

\newcommand{\sigmacrit}{$\Sigma_{crit}$}

\newcommand{\photoz}{photo-z}

\newcommand{\lstarlim}{$0.4 L_*$}
\newcommand{\lvir}{$L_{200}$}

\newcommand{\nvir}{$N_{200}$}
\newcommand{\rvirgal}{$r_{200}^{gals}$}

\newcommand{\ngal}{$N_{gal}$}
\newcommand{\maxbcg}{MaxBCG}
\newcommand{\numNgalBins}{12}
\newcommand{\numLumBins}{16}

\newcommand{\photo}{\texttt{PHOTO}}

\newcommand{\lenszmax}{0.3}
\newcommand{\lenszmin}{0.05}

\newcommand{\photoversion}{\texttt{v5\_4}}

\newcommand{\numTenMpc}{132,473}

\newcommand{\numsource}{27,912,891}

\shortauthors{Sheldon et al.}
\shorttitle{SDSS Cluster Lensing}

\begin{document}

%\title{Ensemble Cluster Lensing in the SDSS I: Measurements}
\title{Cross-correlation Weak Lensing of SDSS Galaxy Clusters I: Measurements}

\author{
Erin S. Sheldon,\altaffilmark{1}
David E. Johnston,\altaffilmark{2,3}
Ryan Scranton,\altaffilmark{4}
Benjamin P. Koester,\altaffilmark{5,6}
Timothy A. McKay,\altaffilmark{7,8,9}
Hiroaki Oyaizu,\altaffilmark{5,6}
Carlos Cunha,\altaffilmark{5,6}
Marcos Lima,\altaffilmark{5,10}
Huan Lin,\altaffilmark{11}
Joshua A. Frieman,\altaffilmark{5,6,11}
Risa H. Wechsler,\altaffilmark{12}
James Annis,\altaffilmark{11}
Rachel Mandelbaum,\altaffilmark{13}
Neta A. Bahcall,\altaffilmark{14}
and Masataka Fukugita\altaffilmark{15}
}
\altaffiltext{1}{Center for Cosmology and Particle Physics, Department of Physics, New York University, 4 Washington Place, New York, NY 10003.}
\altaffiltext{2}{Department of Astronomy, 105-24, California Institute of Technology, 1201 East California Boulevard, Pasadena, CA 91125.}
\altaffiltext{3}{Jet Propulsion Laboratory 4800 Oak Grove Drive, Pasadena ,CA, 91109}
\altaffiltext{4}{Department of Physics and Astronomy, University of Pittsburgh, 3941 O'Hara Street, Pittsburgh, PA 15260.}
\altaffiltext{5}{Kavli Institute for Cosmological Physics, The University of Chicago, 5640 South Ellis Avenue Chicago, IL  60637}
\altaffiltext{6}{Department of Astronomy and Astrophysics, The University of Chicago, 5640 South Ellis Avenue, Chicago, IL 60637.}
\altaffiltext{7}{Department of Physics, University of Michigan, 500 East University, Ann Arbor, MI 48109-1120.}
\altaffiltext{8}{Department of Astronomy, University of Michigan, 500 Church St., Ann Arbor, MI 48109-1042}
\altaffiltext{9}{Michigan Center for Theoretical Physics, University of Michigan, 500 Church St., Ann Arbor, MI 48109-1042}
\altaffiltext{10}{Department of Physics, The University of Chicago, 5640  South Ellis Avenue, Chicago IL 60637}
\altaffiltext{11}{Fermi National Accelerator Laboratory, P.O. Box 500, Batavia, IL 60510.}
\altaffiltext{12}{Kavli Institute for Particle Astrophysics \& Cosmology, Physics Department, and Stanford Linear Accelerator Center, Stanford University, Stanford, CA 94305}
\altaffiltext{13}{Department of Physics, Jadwin Hall, Princeton University, Princeton, NJ 08544}
\altaffiltext{14}{Princeton University Observatory, Peyton Hall, Princeton, NJ 08544.}
\altaffiltext{15}{Institute for Cosmic Ray Research, University of Tokyo, 5-1-5 Kashiwa, Kashiwa City, Chiba 277-8582, Japan.}

\begin{comment}
%for astro-ph
Erin S. Sheldon,
David E. Johnston,
Ryan Scranton,
Ben P. Koester,
Timothy A. McKay,
Hiroaki Oyaizu,
Carlos Cunha,
Marcos Lima,
Huan Lin,
Joshua A. Frieman,
Risa H. Wechsler,
James Annis,
Rachel Mandelbaum,
Neta A. Bahcall,
and Masataka Fukugita
\end{comment}

\begin{abstract}

This is the first in a series of papers on the weak lensing effect caused by
clusters of galaxies in Sloan Digital Sky Survey.  The photometrically selected
cluster sample, known as \maxbcg, includes $\sim$130,000 objects between
redshift 0.1 and 0.3, ranging in size from small groups to massive clusters.
We split the clusters into bins of richness and luminosity and stack the
surface density contrast to produce mean radial profiles. The mean profiles are
detected over a range of scales, from the inner halo (25 kpc/h) well into the
surrounding large scale structure (30 Mpc/h), with a significance of 15 to 20
in each bin.   The signal over this large range of scales is best interpreted
in terms of the cluster-mass cross-correlation function.  We pay careful
attention to sources of systematic error, correcting for them where possible.
The resulting signals are calibrated to the $\sim$10\% level, with the dominant
remaining uncertainty being the redshift distribution of the background
sources. We find that the profiles scale strongly with richness and luminosity.
We find the signal within a given richness bin depends upon luminosity,
suggesting that luminosity is more closely correlated with mass than galaxy
counts.  We split the samples by redshift but detect no significant evolution.
The profiles are not well described by power laws. In a subsequent series of
papers we invert the profiles to three-dimensional mass profiles, show that
they are well fit by a halo model description, measure mass-to-light ratios and
provide a cosmological interpretation.

\end{abstract}


\section{Introduction} \label{sec:intro}

The cold dark matter model (CDM) of structure formation makes a number of
predictions about galaxy clusters which are testable observationally. For
example; the dark matter in clusters should be more evenly distributed than the
observed stars, which are predominantly concentrated in exponential disks and
spheroids.  The radial distribution of the mass in clusters should, in the
mean, follow a predicted universal profile which is a running power law
\citep*[e.g.][]{nfw97}. The number of sub-halos (the centers of which may
correspond to galaxies), and the scatter in that number, are predictable as a
function of halo mass \citep{Kravtsov04a}.  One analytic description of these
sub-halo distributions is known as the halo occupation distribution (HOD).  The
fraction of a cluster's virial mass contained in sub-halos is 10-30\% depending
on how it is counted \citep{MaoFsub04,Gao04}. The number density of halos as a
function of mass has a well-defined form
\citep{PressSchechter74,ShethTormen99,Jenkins01}.  The cluster two-point
correlation and the cluster-mass cross-correlation function are also
predictable in the CDM framework \citep{MoWhite96,SeljakWarren04}; on large
scales, they are proportional to the auto-correlation function of the mass,
which is given by linear perturbation theory.  Each of these predictions
depends more or less on the underlying cosmological model.

In the real world it is difficult to observationally select clusters based upon
their mass. Instead we select clusters based upon some observable and try to
relate that observable to mass.  For example, a HOD-type description, which
uses sub-halos as the basic constituent, can be tested  by counting the number
(and the dispersion in the number) of constituent galaxies as a function of
cluster mass.  The key difficulties are estimating accurate cluster masses and
relating sub-halos to galaxies.  One traditional measure of mass treats the
galaxies as tracers of the gravitational potential.  For example, in a galaxy
redshift survey, one can measure cluster velocity dispersions and, with
additional assumptions about the velocity distribution of the galaxies relative
to the dark matter, infer cluster virial masses \citep[e.g.][]{Carlberg96}.
One can attempt to relate sub-halos to galaxies by matching the observed
abundance of galaxies or clusters to the predicted  abundance of halos or
sub-halos.

With X-ray data, we can use the density and temperature structure of the
baryonic gas, with physical assumptions about the state of the system, to infer
cluster masses \citep{HenryEMSS92,Ebeling98,Bohringer01a}. In this case, one
could in principle bypass the galaxies entirely and use X-ray data alone to
infer, for example, the mass function \citep{ReiprichBohringer02} or the
large-scale clustering \citep{Collins00} of X--ray emitters as a function of
their mean mass.

While useful, dynamical techniques such as galaxy velocity dispersions and
X-ray gas measurements provide limited information on cluster mass profiles,
especially on large scales. Galaxy velocity measurements provide only sparse
sampling of the cluster potential, although stacking velocity measurements from
multiple clusters can yield more precise statistical mass profiles
\citep{Mahdavi04,Katgert04,BeckerVelocities07}. Moreover, both velocity and
X-ray inferences of cluster masses assume that clusters are dynamically simple
systems; this assumption may not be justified in general at the requisite level
of precision, and there are certainly well-known exceptions to it, such as Coma
\citep{Neumann03}.  On scales close to and certainly beyond the virial radius,
one expects the assumption of dynamical equilibrium to break down, requiring
alternative techniques to estimate the associated mass.

Weak gravitational lensing is, in principle, well-suited for studying mass
profiles.  The first weak lensing detection was in a cluster \citep{Tyson90},
and the field blossomed rapidly
\citep{Fahlman94,TysonFischer95,Luppino97,Fischer97,Hoekstra98,
Joffre00,Clowe00,Dahle02,Wittman03,Umetsu05,CloweDMProof06}.  The lensing
effect is sensitive to all mass associated with the cluster, and the
interpretation of the shear in terms of mass is independent of the dynamical
state of that mass. 
%The shear at any point near the line of sight to a cluster, measured from the
%shapes of background sources, is related to the underlying mass distribution.  
If the data permit, mass measurements may be extended to very large scales,
well beyond the virial radius. However, except in the rare cases where the
lensing is very strong, the shape of an individual galaxy gives a very
imprecise measurement of the shear, due to the large variety of intrinsic
source galaxy shapes.  Instead, the shapes of many sources are averaged to
increase the sensitivity. 

For weak lensing measurements, there are additional sources of error that are
not predictable or measurable on a cluster-by-cluster basis. Lensing due to
structures along the line of sight to the source galaxies, such as voids or
distant clusters, can swamp the statistical measurement error
\citep{White02,Hoekstra03}.  Secondly, clusters are spatially correlated with
other clusters, groups, and galaxies. These associated structures boost the
measured lensing signal \citep[e.g.][]{Metzler01,White02}.  This increased
signal can be significant at almost any point in an individual cluster, and is
generally dominant for the average cluster at scales larger than a few virial
radii.  It is difficult, in general, to identify and model these effects for an
individual cluster, which makes recovery of the bound mass uncertain.

The approach we use in this work is a compromise: we average, or ``stack'', the
lensing signal from an ensemble of clusters.  In doing so we cannot recover
detailed information about each cluster.  This is not a significant sacrifice
because the information for individual clusters is not recoverable with high
precision due to the sources of error mentioned above. The gains from this
technique, however, are significant.  The noise due to distant structures along
the line of sight, uncorrelated with the lensing cluster, is negligible in the
mean under the assumption that the universe is homogeneous and isotropic.  The
statistical signal from correlated nearby structures, on the other hand, can be
modeled using CDM; it dominates on large scales but is typically small within
the virial radius.

Another advantage of stacking is that the average cluster mass profile, in the
absence of significant selection effects, must be smooth and spherically
symmetric if the universe is homogeneous and isotropic. In this case the
lensing measurement, which is related to density in a non-local way, can be
inverted directly to the average three-dimensional mass profile modulo the mean
density of the universe \citep{JohnstonInvert07}.

The mean cluster mass profile is best interpreted as the cross-correlation
function between clusters and mass. On small scales this is most sensitive to
the mean density profile of the cluster dark matter halos, while on large
scales it essentially measures how clusters are correlated with the large scale
structure. For a large enough dynamic range in scale these measurements
directly connect the well understood linear growth of perturbations on large
scales to the non-linear collapse of dark matter halos on smaller scales.
These ideas have been discussed and verified in simulations in
\citet{JohnstonInvert07}, and used to reconstruct the galaxy-mass correlation
function in \citet{Sheldon04}.

There is a small literature on ensemble group and cluster lensing.  These
studies have focused mainly on mass-to-light ratios and cosmology
\citep{Hoekstra01c,Parker05}.  In the \citet{Sheldon01} pilot study, we studied
42 SDSS clusters matched to X--ray sources in the Rosat All Sky Survey, and
demonstrated the feasibility of ensemble cluster lensing in the Sloan Digital
Sky Survey (SDSS).  

This work is the first in a series of papers on statistical cluster lensing in
SDSS. The clusters used in this study are drawn from a superset of the
recently released \maxbcg\ catalog of
\citet{KoesterAlgorithm07,KoesterCatalog07}, extending that catalog to lower
richness objects.  We present lensing measurements in bins of cluster richness
and cluster $i$-band luminosity, detailed descriptions of our methods, and
tests and corrections for systematics. We also present some basic statistics
about these profiles, such as tests for redshift dependent signals and
comparisons with power law models (which are not a good description).  The
idea is to present basic and stable results that will not depend on models of
the moment or assumptions about the selection function.  This paper is
accompanied by \citet{JohnstonLensing07} in which we present detailed analysis
and modeling of the profiles, such as non-parametric inversions of the lensing
profiles to three dimensional density and aperture mass.  In that paper we
also model these profiles to extract cluster virial masses and concentrations,
large scale cluster-mass cross correlations and bias, and the mass-observable
relations such as $M-L$ and $M-N_{gals}$.  This paper is also accompanied by
\citet{SheldonM2L07}, which focuses on cluster mass-to-light ratios.  A paper
on the mass function \citep{RozoMassFunction07} is forthcoming.

We assume the universe is described by a Friedman-Robertson-Walker cosmology
with $\Omega_M$ = 0.27, $\Omega_{\Lambda}$ = 0.73, and H$_0$ = 100 $h$
km/s/Mpc.  All distances are measured in physical, or proper, coordinates, not
comoving.  The basic measures of richness and luminosity we will refer to as
\nvir\ and \lvir.  These are the counts and i-band luminosity for galaxies with
$L_i > 0.4 L_*$, colors consistent with the cluster ridge-line, and projected
separation less than \rvirgal. See \S \ref{sec:clustersel} for more details.

\section{Lensing Formalism and Inversions} \label{sec:lensmeas}

This section includes a brief description of the lensing formalism used in this
paper. More details can be found in \citet{Sheldon04}.

Gravitational lensing is the apparent bending of light as it passes massive
objects.  The actual path of light is not generally observable, but the
distortions produced in the images of sources are.  Any distortion produces
correlations in the shapes and orientations of background sources, and these
correlations are measurable.  Note, if some fraction of the sources are in
fact not in the background but associated with the lens and if there are
``intrinsic alignments'' between galaxies in or near the lensing cluster, then
the lensing signal will be contaminated.  We address this issue in more detail
in \S \ref{sec:systematics}.

For statistical weak lensing measurements, the basic observable is the tangential
shear \gammat\ which, for small shears, is simply proportional to the change in
shape of the galaxy
\begin{equation} \label{eq:induced_shear}
e_+ = 2 \gamma_T \mathcal{R} + e_+^{int} ~,
\end{equation}
where $e_+$ is the measured ellipticity of the galaxy, in the tangential frame
of reference, and $e_+^{int}$ is the intrinsic shape of the galaxy; the
quantity $\mathcal{R}$ is the ``responsivity''or ``shear polarizability'' of
the galaxy. It encodes how strongly the image responds to an applied shear, and
is measurable in the mean from the ensemble of galaxy shapes \citep{Bern02}.
The intrinsic shape $e_+^{int}$ is the primary source of noise, and its
properly weighted RMS value is known as the ``shape noise'', $\sigma_{SN}^{2} =
\langle (e_{+}^{int})^{2}\rangle$.  In the absence of intrinsic alignments,
$e_+^{int}$ has zero mean.

The azimuthally averaged tangential shear is related to the geometry
of the lens-source-observer system and the projected mass density of the lens:
\begin{equation} \label{eq:gammat}
\gamma_T (R) \times \Sigma_{crit} = \bar{\Sigma}(< R)
-\bar{\Sigma}(R) \equiv \Delta \Sigma ~,
\end{equation}
where $\bar{\Sigma}(<R)$ is the mean projected mass density within the disk of
transverse radius $R$, and $\bar{\Sigma}(R)$ is the mean within the annulus
used to measure the shear.  The proportionality, $\Sigma_{crit}$, encodes the
geometry of the lens-source-observer system:
\begin{equation} \label{eq:sigmacrit}
\Sigma_{crit}^{-1} = \frac{4 \pi G D_{LS} D_L}{c^2 D_S}~,
\end{equation}
where the $D_{j}$ are the angular diameter distances to lens, source, and
between lens and source.  Note, the shear measured at 45 degrees relative to
the tangential, \gammacross, should be zero if the signal is due to lensing.
As will be discussed in later sections, we have excellent photometric redshifts
for each of the cluster lenses, as well as photometric redshifts for each
source galaxy; together these provide an estimate of \sigmacrit\ and
allow us to convert the tangential shear to a measurement of \deltasig.

The signal to noise ratio in \deltasig\ for a typical lens in the SDSS is much
less than unity due to the low redshift of the lenses and sources and the
relatively low number density of the source catalog. To increase the
sensitivity, we average the \deltasig\ measurements from an ensemble of lenses
of similar optical properties.  This mean signal, measured as a function of
projected separation, $R$, is related to the cross-correlation between the
lenses and the density field.  Under the assumption of statistical isotropy
(after stacking, the lenses look spherically symmetric), the mean \deltasig\
profile can be inverted to the three-dimensional excess density profile:
\begin{eqnarray} \label{eq:invert}
-\frac{d\Sigma}{dR} & = & \frac{d\Delta\Sigma}{dR} + 2\frac{\Delta\Sigma}{R}
           \nonumber \\
\Delta \rho(r) & \equiv & \rho(r) - \bar{\rho} = \frac{1}{\pi} \int_r^{\infty} dR \frac{-d\Sigma/dR}{\sqrt{R^2-r^2}} 
\end{eqnarray}
where \deltarho\ is the mean excess density, relative to the mean density of
the universe $\bar{\rho}$.  The mean density of the universe will not
contribute to the shear in equation \ref{eq:gammat}, and thus does not 
contribute to \deltasig\ or \deltarho.  This inversion technique
was used in \citet{Sheldon04}, and the details are presented in full in
\citet{JohnstonInvert07}.  The integral in equation
\ref{eq:invert} cannot be taken to infinity; in practice this limits the useful
range of radii to about 2/3 of the largest radius where measurements are
available. As shown in \citet{JohnstonInvert07}, the total mass within radius
$r$ can also be recovered from \deltarho\ and \deltasig.
%This includes the mass
%interior to the minimum radius, due to the non-locality of \deltasig\ (see
%equation \ref{eq:gammat}).  Essentially, $\Delta\Sigma(R) \times \pi R_{min}^2$
%gives the cylindrical mass within projected radius $R_{min}$, which can be
%inverted assuming spherical symmetry. The integral of equation
%\ref{eq:invert} will give mass between $R_{min}$ and $R_{max}$.

The assumption introduced in equations \ref{eq:invert} is that the correlation
function is statistically isotropic. This follows from the isotropy of the
universe as long as the cluster finder does not introduce a preferred
direction.   For example, if the \maxbcg\ cluster finder, described in \S
\ref{sec:clustersel}, preferentially chose structures oriented along the line
of sight this would violate the assumption that the correlation function is
isotropic.  Tests and predictions of this effect will be presented in
\citet{JohnstonLensing07}.

\section{Data} \label{sec:data}

\begin{figure*}[t]
\plotone{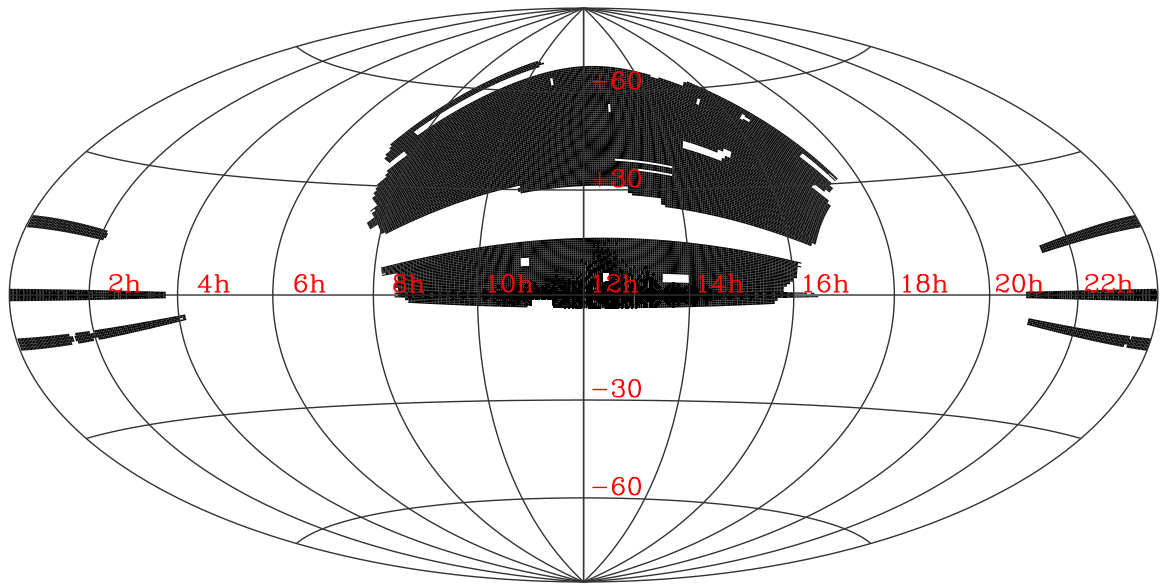} \figcaption{Hammer-Aitoff projection
  of the sky, with the area used for this analysis shaded in black.  This is a
  subset of the SDSS DR4. \label{fig:aitoff}}
\end{figure*}

The galaxies used for cluster selection and shear measurement were drawn from
the Sloan Digital Sky Survey \citep[SDSS;][]{York00}. The area used for this
study is somewhat smaller than the SDSS data release 4 \citep[hereafter
DR4]{edr,dr4}.  The coverage area is shown in Figure \ref{fig:aitoff}.

The SDSS observing mode is time-delay-and-integrate, with the camera reading
out at the scan rate, resulting in an effective exposure time of 54
seconds. The camera layout, described in \citet{Gunn98,Gunn06}, comprises 6
columns of 5 CCDs. Each CCD in a column is covered with one of the 5 SDSS
filters \citep{Fukugita96} and objects pass through each different filter in
turn, resulting in nearly simultaneous imaging.  The gaps between columns are
about a CCD width, and the resulting gaps in the imaging data is scanned on an
alternate night.

These data are reduced to object lists through a series of calibration
\citep{Hogg01,Smith02,Tucker06}, astrometric \citep{Pier03}, and photometric
\citep[\photo]{LuptonADASS01} pipelines.  The version of \photo\ used for these
data is \photoversion.  Among the parameters for each object are the position
(RA,DEC), various fluxes, and moments of the light distribution used for shear
estimation.  These same properties for the PSF, measured from bright stars, are
interpolated across the image and used for shape corrections (\S
\ref{sec:corrections}).  Only data that pass a series of automated quality
assurance tests are included in the final catalog \citep{Ivezic04}.

\subsection{Masks} \label{sec:data:masks}

The area used for this study (see Figure \ref{fig:aitoff}) was characterized
using the SDSSPix pixelization scheme
\footnote{http://lahmu.phyast.pitt.edu/$\sim$scranton/SDSSPix/} to define a
window, or mask of the available area.  This same scheme has been used for
other clustering analyses \citep{ScrantonISWAstroPh03, ScrantonMag05} and
lensing analyses \citep{Sheldon04}.  This mask includes survey boundaries and
holes in the survey area. We also exclude regions with inferred extinction
greater than 0.2 mag in $r$ (according to the \citet{Schlegel98} dust maps)
from the analysis.  Only objects that pass this mask were included in the
sample; this includes both the lensing source galaxies and the clusters. This
is the area shown in Figure \ref{fig:aitoff}. This mask was also used for the
edge cuts described in \S \ref{sec:clustersel}.

\subsection{Source Galaxy Selection} \label{sec:sources}

Candidate source galaxies were drawn from the available area described in \S
\ref{sec:data:masks}. Stars were separated from galaxies using the Bayesian
method described in \citet{Scranton02} and \citet{Sheldon04}.  In addition,
only galaxies whose size was measured to be much larger than the PSF were
used. The cut used is the same as that used in \citet{Mandelbaum06}: the
resolution parameter $R$, which is roughly unity minus the square of the ratio
of object PSF size to object size, must be greater than 1/3.  The combination
of these two cuts is quite conservative: tests based on comparisons with the
deeper co-added southern stripe indicate the stellar contamination is less than 1\%.

\begin{figure}[t]
%\epsscale{\epssmall}
\plotone{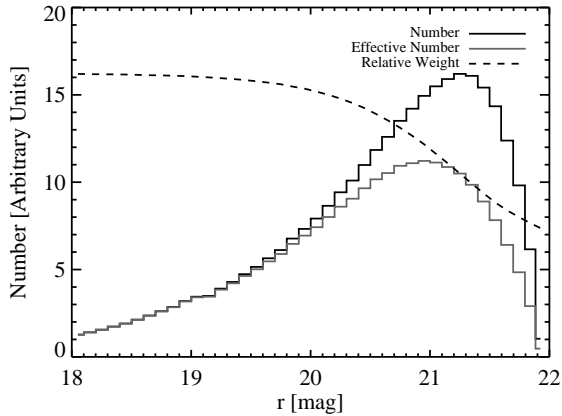} 
\caption{Distribution of source galaxy magnitudes after applying the cuts
  described in \S \ref{sec:sources}.  The black line is the normalized
  histogram and the gray line is the effective number, normalized to the bright
  end, after including the weight. The relative weight, shown in arbitrary
  units as the dashed curve, is an inverse shear variance weight.
    \label{fig:number_counts}}
\end{figure}

We discarded objects for which the shape measurement did not converge, large
measurement errors $\sigma_e > 0.4$ and the very tail of the ellipticity
distribution $e > 4$ (the ellipticity can exceed unity after the dilution
correction).  Only objects with detections in all five SDSS bandpasses were
used, since the accuracy of photometric redshifts is significantly reduced
otherwise.  Finally, only objects brighter than $r = 22$ were included.  The
final catalog contains \numsource\ galaxies.  The distribution of $r$
magnitudes is shown in Figure \ref{fig:number_counts}.  The gray curve shows
the weighted, ``effective'' number, where the weight is the inverse shear
variance, $1/(\sigma_e^2 + \sigma_{SN}^2)$. Note there are a fair fraction of
galaxies fainter than 21.5, but these get little weight in the final analysis.

\subsection{Shape Measurement and Correction} \label{sec:data:shapes}

The details of the shape measurement were given in \citet{Sheldon04}, and are
an implementation of the techniques presented in \citet{Bern02}.  With this
method, the moments of an elliptical Gaussian weight function are matched to
those of the object in question through an iterative algorithm.  We measured
second and fourth order moments for all objects in the survey using this
method; these parameters are in the SDSS database.

The second order moments $Q_{m,n}$ for each object are combined into the shape
parameters,
\begin{eqnarray} \label{sec:data:eq:e1e2}
Q_{m,n} & = & \sum_{m,n} I_{m,n} W_{m,n} x_m x_n \nonumber \\
e_1 =  { {Q_{1,1} - Q_{2,2}} \over {Q_{1,1} + Q_{2,2}} }, & &
e_2 = { {2Q_{1,2}} \over {Q_{1,1} + Q_{2,2}} },
\end{eqnarray} 
where $I_{m,n}$ is the intensity at pixels $m,n$, $W_{m,n}$ is the radial
weight function, and $x_m$ are the pixel coordinates relative to the centroid
of the galaxy light.  These shape parameters are used directly in the shear
estimation, as shown in equation \ref{eq:induced_shear}. The fourth order
moment accounts for the non-gaussianity of the object and is part of the
resolution parameter, or smear polarizability \citep{fis00,Bern02}.

The point spread function (PSF) smears and changes the shape of galaxies, which
can be mistaken for the effects of lensing.  In order to correct for these
effects, the PSF was modeled from bright stars and interpolated to the position
of each galaxy using a KL decomposition. This algorithm was described in detail
by \citet{LuptonADASS01} and \citet{Sheldon04}.  The effect of the PSF was then
corrected for using the techniques of \citet{Bern02} combined with the
``re-Gaussianization'' method of \citet{HirataCalib03}.  This method treats the
PSF as a Gaussian plus a small non-Gaussian component, which is generally a
good representation of the SDSS PSF. A convolution kernel is used to transform
the image, removing the effect of the non-Gaussian component.  The remaining
Gaussian PSF component can be corrected for exactly using the formulas in
\citet{Bern02}.

These corrections for PSF convolution can be thought of as two separate
corrections: the first corrects for the smearing by the finite PSF which makes
the object look more round and affects the shear calibration.  This is often
referred to as PSF dilution.  The second alters the shape of the galaxies due to
the anisotropy of the PSF. These are often referred to as ``multiplicative''
and ``additive'' biases, respectively.  In practice these two biases are
coupled.

%The shear calibration errors for this sample are essentially at the same level
%as those in the study of \citet{Mandelbaum06}, and should be less than 10\% (95\%
%confidence) for the majority of the source sample ($r < 21$) and better than
%20\% for fainter galaxies ($r > 21$). The additive errors can be corrected
%for by looking at the lensing signal around random points, which should
%be zero in the absence of systematics.  This correction is addressed in
%\S \ref{sec:corrections}.

\subsection{Photometric Redshifts} \label{sec:photoz}

The photometric redshifts (\photoz) were calculated using a Neural Network
based on the training set method of \citet{CollisterLahav04}. The
spectroscopic training set was collated from the SDSS spectroscopic survey and
various surveys with coverage overlapping the SDSS. From a total of
$\sim$60,000 galaxies with measured redshifts, $\sim$45,000 were from the SDSS
main galaxy sample \citep[roughly $r < 17.6$, see][]{Strauss02}, $\sim$14,000
were from the SDSS luminous red galaxy (LRG) sample \citep{Eisenstein01},
$\sim$1,500 were from the CNOC2 survey \citep{Yee01}, and $\sim$300 were from
the CFRS survey \citep{Lilly95}.  The training set covers a redshift range
between 0 and 1.  Of the $\sim$60,000 objects, 30,000 were used to train an
artificial neural network while the remaining objects (the validation set)
were used to check the results of the trained network.  The resulting RMS
scatter in photo-z's for the validation set was $\sim$ 0.04 over the entire
magnitude range, with more scatter at the faint end.  Recall that galaxies
with fainter magnitudes get less weight as shown in figure
\ref{fig:number_counts}.  The details of these methods can be found in
\citet{Oyaizu07}. 

The resulting photometric redshift sample has remaining errors, and these
errors are a function of redshift.  We used the validation set to constrain the
true redshift distribution in a given bin of photoz.  In order to properly
sample the true distribution of the photometric sample, the spectroscopic
sample was weighted such that the probability distributions of each of the five
SDSS bandpasses matched that of the photometric sample.  We then used these
weights to produce a weighted histogram of spectroscopic redshifts for the
given photoz bin.  More details about this weighting scheme can be found in
\citet{Lima08}.  Those authors found that this method reproduced the underlying
distributions with a high level of accuracy in simulations where the only
variables for selection were magnitude.  In the real data, other factors may be
important such as surface brightness and angular size.  These factors have not
yet been addressed and may lead to additional errors.

\begin{figure}[t]
\plotone{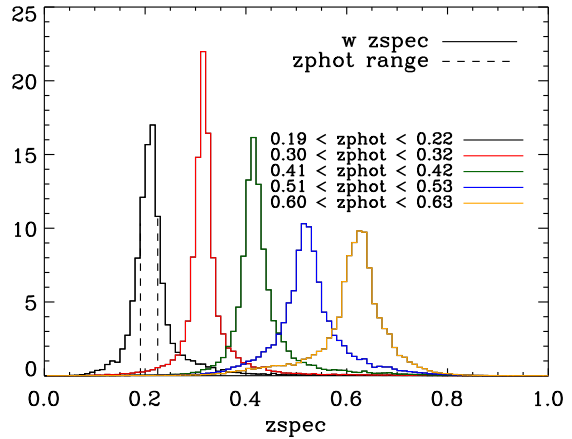}
\figcaption{ Weighted distribution of spectroscopic redshifts in a few bins of
photometric redshifts.  We chose weights such that the weighted distributions
of each of the five SDSS bandpasses match that of the photometric sample in the
given bin.  The error bars come from bootstrap re-sampling the spectroscopic
validation set.
\label{fig:wzdist}
}
\end{figure}

Example distributions for a few photoz bins are shown in figure
\ref{fig:wzdist}.  The peak of the distributions is relatively unbiased,
indicating the neural network tends to find the maximum likelihood.  However,
the distributions are broad and even skewed in some cases. 

The lensing \deltasig\ depends on the distances to lens and source through the
inverse critical density.  For each lens-source pair, we integrated over the
above distributions in redshift to get the expected inverse critical density.
The formalism we used was the same as \citet{Sheldon04} except here we
integrated over the full distribution determined as outlined above, whereas in
that work we used the error estimates that came out of a chi-squared analysis
over galaxy templates.  In this work we applied no additional prior to the
distribution.  In the end the results using only these distributions differed
from calculations treating the photozs as perfect by only a percent.

We also repeated our analysis for sources in each of the redshift ranges shown
in figure \ref{fig:wzdist}, as well as for relative redshift thresholds
$z_{source} > z_{lens} + \delta z$ up to $\delta_z=0.2$.  Although the results
are generally noisier for the subsamples, we saw consistent results for each
sample.

\subsection{Cluster Selection} \label{sec:clustersel}

The \maxbcg\ algorithm \citep{KoesterAlgorithm07,KoesterCatalog07} was used to
identify clusters in SDSS imaging data using three observational properties of
rich galaxy clusters.  These properties are 1) spatial clustering, 2)
clustering in color-space (the red sequence), and 3) the presence of a
brightest cluster galaxy (BCG) spatially coincident with the approximate
center of the cluster.  These model components are folded into a
redshift-dependent likelihood function, which has built-in predictions for the
colors of the red-sequence, the color-magnitude properties of the BCG, and the
spatial distribution of cluster members \citep*[NFW,][]{nfw97}.  Every object
in the input galaxy catalog is then evaluated for its likelihood of being a
BCG in an over-density of bright ($>$\lstarlim\ in i-band) red galaxies
(within $\pm 2\sigma$ of the red sequence) at a grid of redshifts.  The
$\sigma$ in this case is the intrinsic width of the red sequence which is 0.05
in $g-r$, plus the measurment error for each object.  A maximum likelihood
redshift is determined for each object, and these potential BCGs are ranked by
decreasing maximum likelihood.  In a manner analogous to the spherical
over-density algorithm employed in N-body simulations, the highest likelihood
potential BCG in the survey is deemed a BCG at the center of a cluster. Any
lower likelihood objects in the ranked list within a projected \rvirgal\ (see
below) and $\pm 0.02$ in redshift of this first BCG are eliminated as possible
cluster centers.  The process is repeated for the next object in the ranked
list, given that it is not in the exclusion region of the first object. This
prescription is iteratively applied to the entire list, thereby generating the
final cluster catalog.
  
The initial parameters in the cluster catalog include a maximum likelihood
redshift (between \lenszmin\ and \lenszmax) and an initial richness estimate,
\ngal\ (the number of red sequence galaxies within 1 Mpc).  The sample is well
understood in terms of completeness, purity, and \photoz\ accuracy within the
redshift range $[0.1,0.3]$; at lower redshifts the \photoz\ is less reliable so
these objects are discarded. Over the redshift interval $[0.1,0.3]$, the
\photoz\ 's have a small bias $\sim 0.004$ which we correct and scatter that is
$\lesssim 0.01$.  The sample appears to have a number density that is roughly
independent of redshift, although the presence of a supercluster at redshift
0.08, the so-called ``Sloan Great Wall'', dominates the statistics at low
redshift \citep{Gott05}.
 
After these initial parameters are determined for each cluster candidate,
measurements of cluster richness and redshift are refined.  A scaled radius
\rvirgal\ is determined from the \ngal\ -- \rvirgal\ relationship measured by
\citet{Hansen05}; it is the radius at which the mean luminosity density reaches
200 times the mean value predicted by the luminosity function.  The number of
galaxies \nvir\ and the i-band luminosity \lvir\ are then
re-counted within \rvirgal\ with the same selection criteria as above.
%Note that \citet{Hansen05}
%was based on an earlier incarnation of the \maxbcg.  We have begun
%re-calibrating this relationship with the new catalog, and preliminary results
%show that the normalization has changed but the scaling has not \citep[in
%prep.]{Masjedi06b}.  Thus the normalization of \nvir\ and \lvir\ will also be
%re-calibrated. Finally, the photometric redshift was remeasured in a continuous
%way to remove the gridding effect.

\begin{figure}
    \plotone{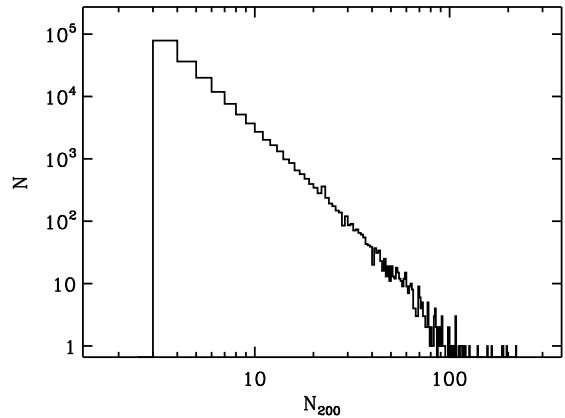}

    \caption{Histogram of \nvir, the number of red galaxies brighter than
    \lstarlim\ within \rvirgal, for the \maxbcg\ cluster sample.
    \label{fig:ngals_function} }

\end{figure}

\begin{figure}
    \plotone{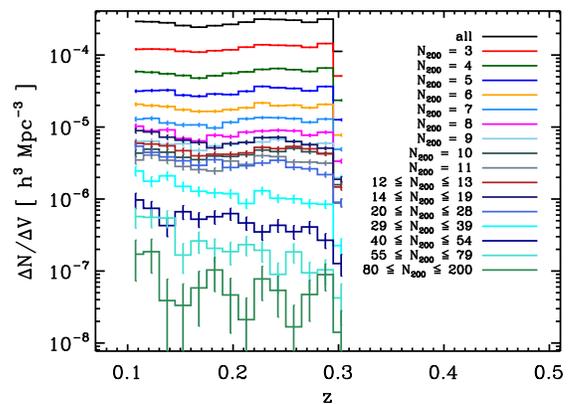}

    \caption{Number density as a function of redshift for differential bins of \nvir.\label{fig:zdensity}}

\end{figure}

Some statistics for the final cluster sample are shown in Figures
\ref{fig:ngals_function} and \ref{fig:zdensity} The \nvir\ function is shown
in the left panel.  We have extended this sample to \nvir=3, substantially
lower than the sample from \citet{KoesterAlgorithm07}.  These objects have
substantially more noise in their \photoz\ and richness measures, but as we
will see in the following sections they have a well-measured lensing signal.
The added noise does, however, complicate the interpretation of the lensing
results.  In paper II \citep{JohnstonLensing07} we interepret the \maxbcg\
selection function using the best available simulations, but a full analysis
will require large volume high resolution simulations that are not currently
available.  Nevertheless, we present the lensing measurements here in
anticipation of future improvements to the simulations.


Figure \ref{fig:zdensity} shows the cluster abundance as a function of
redshift.  The overall sample is close to volume limited, but the
distributions in individual \nvir\ bins have significant features in redshift.
This is due to some combination of the redshift-dependent model for 0.4L*, the
definition of the red-sequence, and true evolution in the number of galaxies
at fixed mass and redshift. It is difficult to disentangle these
contributions. 
%{\bf (XXX Need input from Risa and Ben here with insight from the sims)}.

Geometrical edge cuts were applied before using the clusters for lensing, in
addition to the basic mask cut applied in \S \ref{sec:data:masks}.  As will be
explained in \S \ref{sec:corrections}, we require the searchable area
surrounding each lens center to be either a full disk or a half disk on the
sky. This is to guarantee there are pairs of sources at 90 degree separation
with respect to the lens center to cancel residual PSF systematics.  We checked
each lens against the mask to guarantee this.  The maximum search radius is a
function of redshift, so lenses at low redshift are more likely to hit an edge.
We discarded 21\% of the available clusters leaving \numTenMpc.

\section{Corrections to the Lensing Profiles} \label{sec:corrections}

Two corrections were made after the basic lensing measurement was complete.
The first corrects for the so-called ``additive'' bias outlined in section \S
\ref{sec:data:shapes}, and the second corrects for the clustering of source
galaxies with the lenses.

\subsection{Correction for Additive Bias}

The additive bias is due the residual PSF induced ellipticity, left over from
imperfect PSF interpolation which results in small inaccuracies in the galaxy
shapes after PSF correction.  For the most part, this is a random effect
across the survey that cancels in the average, but because it is correlated
over the scales we are interested in it does not cancel compeletely.  Also,
residuals that are constant across the shear measurement area for a given lens
will cancel as long as there are pairs at 90 degree separation, which prompted
the edge cuts outlined in \S \ref{sec:clustersel}.  Components that are not
effectively constant, however, may be present.
These correlated residuals are confined to an SDSS stripe for the most
part, and are thus correlated on angular scales less than a few degrees.

This effect was checked using random points generated over the same area as the
lenses and sampling the same systematics. We used the mask generated for this
data (see \S \ref{sec:data:masks}), including the edge cuts applied in \S
\ref{sec:clustersel}.  Random points were assigned redshifts such that the
redshift histograms of clusters and randoms are proportional when binned at
$\Delta z = 0.01$.  This histogram matching was performed separately for all
binnings of the clusters shown below; different samples may have different
redshift distributions. This is important because systematics depend mainly
on angular scale, which affects different physical scales as
a function of redshift.

\begin{figure}[b]
\centering
\plotone{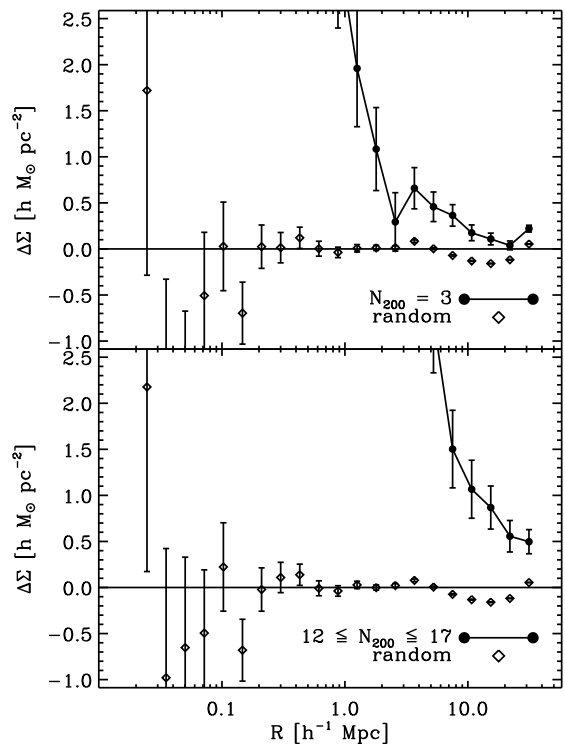}
\caption{Lensing measurement \deltasig\ around random points as compared to
  clusters.  The top panel compares randoms to clusters for the \nvir=3 bin;
  the bottom panel shows the $12 \le $\lvir$ \le 17$ bin.  The non-zero
  detection at large scales is indicative of residual additive systematics in
  the PSF correction. The random signal dominates on large scales for the
  relatively low signal bin shown in the top panel. The contamination is less
  important but still significant for the intermediate bin shown in the bottom
  panel. The random signal is subtracted from the cluster signal for all
  samples presented in this work.  
  \label{fig:randoms30Mpc}}
\end{figure}

The results for two sets of random points are shown in Figure
\ref{fig:randoms30Mpc}.  The top panel is for \nvir$ = 3$, the lowest richness
bin, which has the weakest signal. The bottom panel is for a moderately high
richness bin, $12 \le$\nvir$ \le 17$.  On very large scales, the signal for
the top panel is strongly affected by this additive systematic, while the
signal in the bottom panel is less affected due to the higher signal. The
scale of these corrleations is a few degree for the mean redshift of 0.25, as
expected.  Note, on smaller scales there may be additional correlations that
are not detected due to the higher noise, but they are not relevant as the
signal from the clusters is much higher.  This residual additive bias is
subtracted from the lensing signal for all samples presented below
\footnote{Note that this technique of correcting for residual PSF anisotropy
using random points only works for lens-shear cross-correlations, not for
shear-shear (so-called cosmic shear) correlations.}.

\subsection{Correction for Source-Lens Correlations}  

\begin{figure}
\centering
\plotone{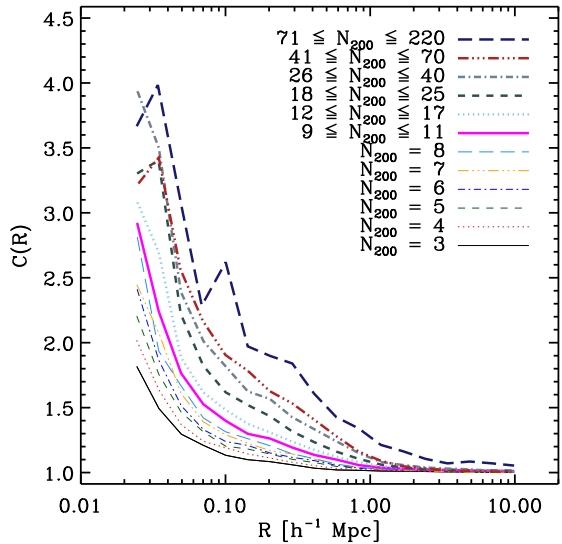}
\caption{Correction factor for the clustering of source galaxies
  with the clusters in bins of \nvir. The multiplicative factor $C(R)$ is
  calculated and applied for all profiles shown 
  herein. \label{fig:corr10Mpc}}
\end{figure}

The second error we correct for is the clustering of source galaxies with the
lensing clusters.  Although photometric redshifts help to remove cluster
members from the source sample, there is still a significant contamination,
especially for the richer clusters. These contribute zero to the shear,
diluting the inferred profile.  Because the fraction of cluster members
polluting the source sample is a function of radius, this contamination alters
the shape of the profile.  This contamination was estimated by computing the
correlation function between clusters and sources \citep{fis00,Sheldon04}. The
sources are weighted exactly as in the lensing measurement. The result is shown
in Figure \ref{fig:corr10Mpc} for 12 bins of the \nvir\ measure. In some bins
the correction factor is quite large at small scales, but is well-understood.

\section{Results} \label{sec:results}

\begin{figure*}[h]
\centering
\plotone{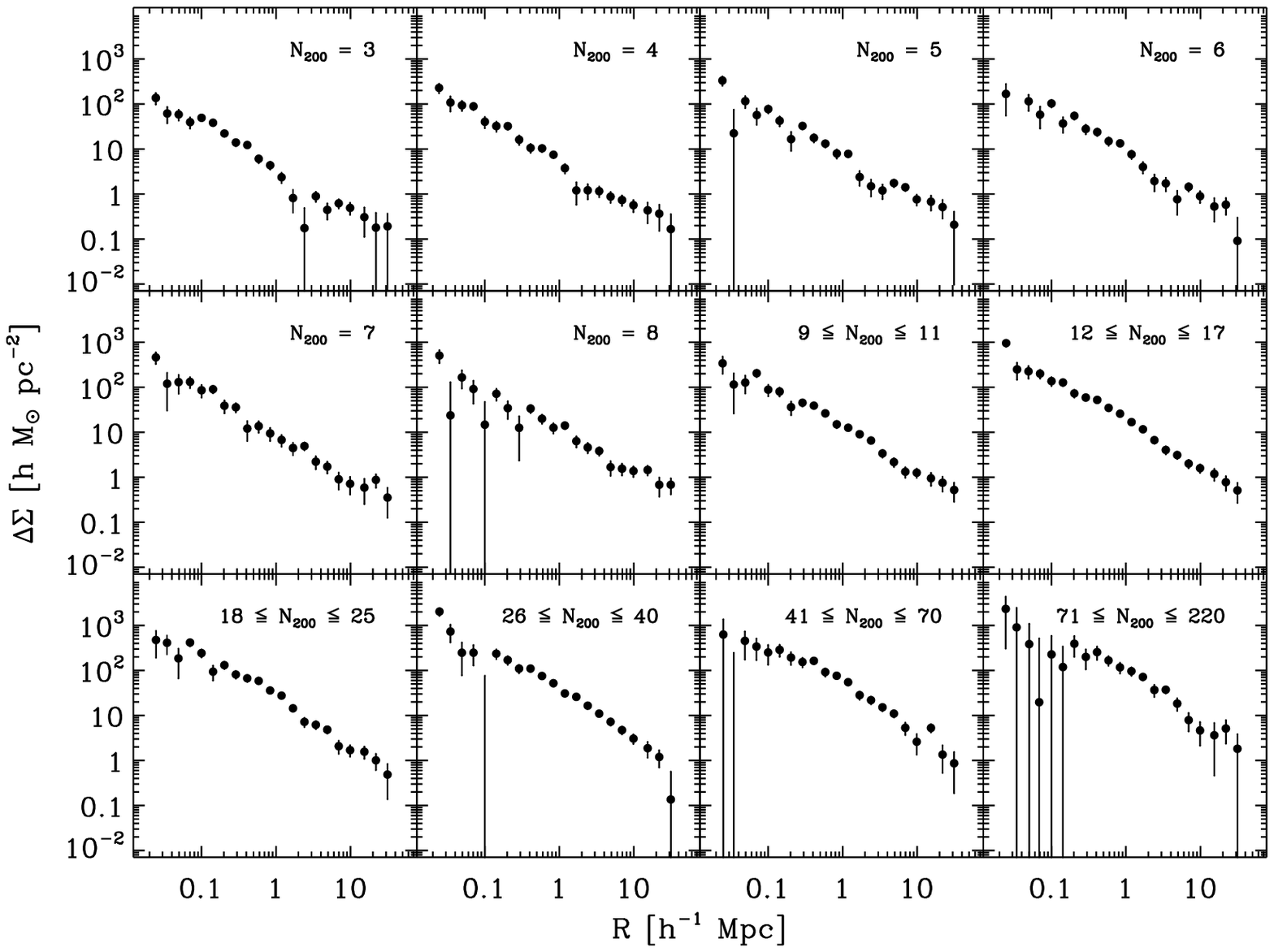}
\caption[\deltasig\ in \nvir\ bins]{
\deltasig\ from 25 to 30$h^{-1}$ Mpc in \numNgalBins\ bins of \nvir, the number
of galaxies $(>0.4 L_*)$ within \rvirgal.  The signal measured around random points is
subtracted from these profiles (see Figure \ref{fig:randoms30Mpc}).  The
correction for clustering of sources with the lenses is also applied (see
Figure \ref{fig:corr10Mpc}).  The errors are from jackknife re-sampling.
\label{fig:deltasig_ngals200_12}
}
\end{figure*}

The lensing \deltasig\ profiles in \numNgalBins\ bins of \nvir, are shown in
Figure \ref{fig:deltasig_ngals200_12}.  There is a significant detection for
each of the bins, with signal-to-noise ratio $\sim 15-20$ for all bins.  The
corrections described in \S \ref{sec:corrections} have been applied to each of
these profiles.  

\epsscale{1.2}
\epsscale{1}

The errors are the diagonal elements of the covariance matrix derived from
jackknife re-sampling.  We use relatively small jackknife regions for this
analysis, about the width of an SDSS stripe.  This choice is made based on
indications that variations in the systematics are the dominant source of
spatial variation in the signal: the PSF also varies with roughly this scale,
and of course the scans are this size.  This is further supported by the fact
that for much larger jackknife regions the errors are consistent with the
standard Gaussian error propagation. For this jackknife scale errors are consistent with
simple error propagation for $R < 1 h^{-1}$ Mpc, but are substantially larger
on $R > 5 h^{-1}$ Mpc scales. The covariance matrix becomes non-diagonal
for $R > 5 h^{-1}$ Mpc, where the first off-diagonal terms are about 30\%.
Note, at the median cluster redshift 0.25, the width of a stripe is roughly
25 $h^{-1}$ Mpc.

The amplitude of the lensing profile is a strong function of \nvir, as expected
if the number of galaxies correlates with the mass of the cluster.  However,
the interpretation of these profiles in terms of halo masses is complicated, as
with any cluster mass measurement, by the contributions from the neighboring
large scale structure.  In fact, for most of the profiles, the signal at
separations larger than two megaparsecs is entirely dominated by associated
large scale structure; this feature allows us to measure halo bias in addition
to halo mass.  The interpretation of these profiles in terms of halo mass and
correlated mass (i.e. the halo model) is presented in detail in the companion
paper \citet{JohnstonLensing07}.

\begin{figure}[t]
\centering
\plotone{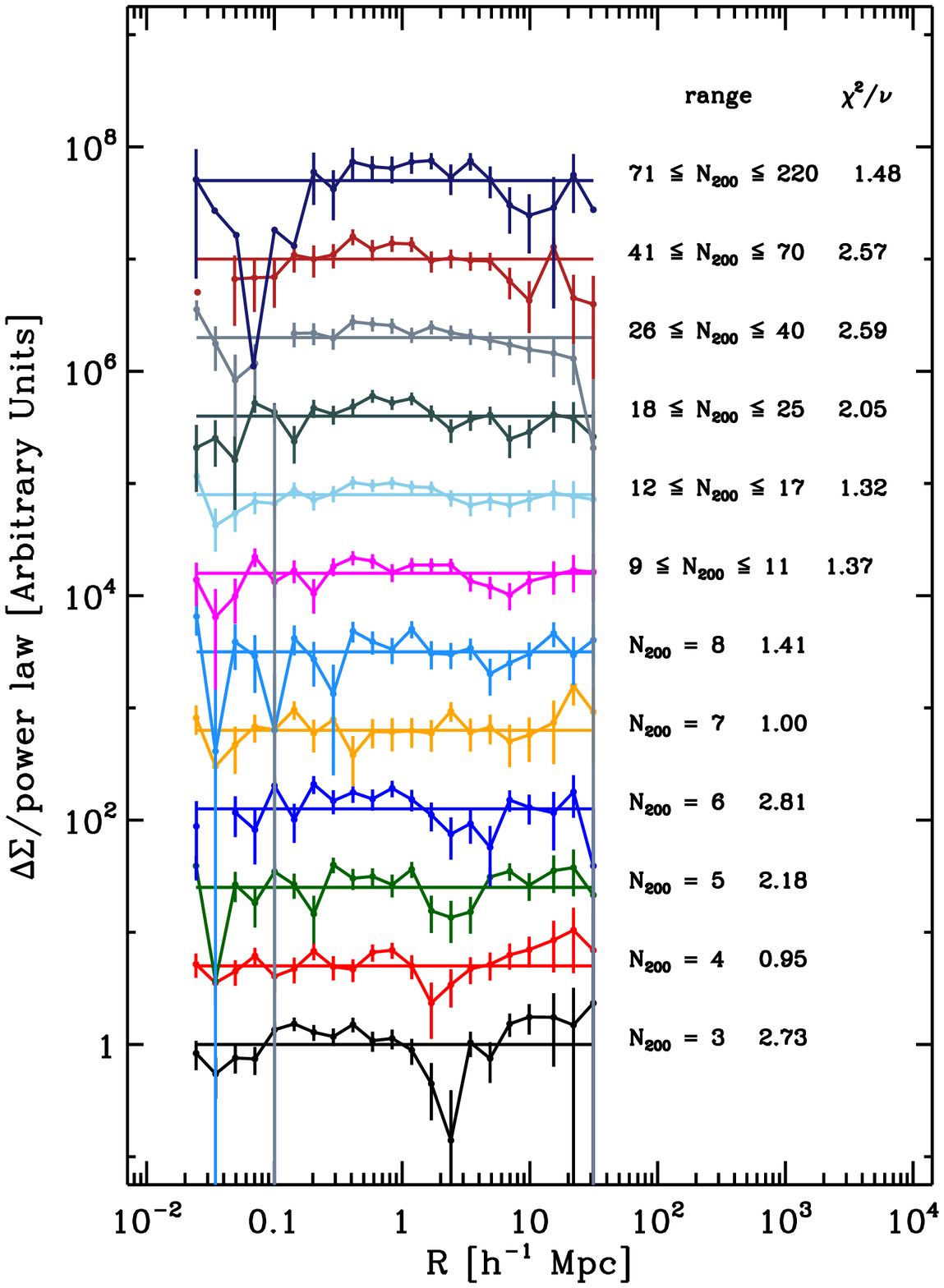}
\caption[]{
  Power law fits in each \nvir\ bin.  For each bin, the best fitting power law
  (see Table \ref{tab:powfits}) has been divided out and the signal scaled
  arbitrarily to separate the profiles visually.  For clarity, the error bars
  for points with S/N $<$ 1 have been suppressed for the top two bins.  Each
  curve is labeled by its richness range and reduced $\chi^2$ for the power law
  fit. The profiles are not generally good fits to a power law and demonstrate
  systematic deviations as a function of scale.
  \label{fig:deltasig10Mpc_ngals200_powfits}}
\end{figure}

A number of the profiles in Figure \ref{fig:deltasig_ngals200_12} show
deviations from a power law.  Power law fits for each bin are shown in Table
\ref{tab:powfits}.  The fits were performed using the full covariance matrix.
All but the 2nd and 5th bins have $\chi^2$ per degree of freedom greater than
1.3 (d.o.f. = 19), indicating a poor fit. This is demonstrated visually in
Figure \ref{fig:deltasig10Mpc_ngals200_powfits}, where in each case the
profile has been divided by the best-fitting power law. The curves
systematically have a shallower logarithmic slope at small radius and a
steeper slope at large radius. The shape on scales $\lesssim 100 h^{-1}$ kpc
may be affected by systematics, as outlined in \S
\ref{sec:systematics:lensmeaserr}.  The fits are dominated by larger scales,
however, and removing these points does not improve the fit significantly.
There is also evidence of an upturn again at the largest radii which may be
interpreted as the ``two-halo term'', the transition to correlations with
neighboring large scale structure.  These features move to larger radius for
higher \nvir.  The shape of these curves is in qualitative agreement with a
model where the inner halo is NFW-like with a transition to linear
correlations on large scales. In \citet{JohnstonLensing07} we demonstrate
quantitatively that such a model is in fact a good fit to this data.

\begin{deluxetable}{cccc}
\tabletypesize{\small}
\tablecaption{Power law fits for \nvir\ Bins \label{tab:powfits}}
\tablewidth{0pt}
\tablehead{
  \colhead{Bin}       &
  \colhead{A}         &
  \colhead{$\alpha$}  &
  \colhead{$\chi^2/\nu$}
}
\
\startdata

$N_{200} = 3$ &        3.2 $\pm$        0.2 &       1.06 $\pm$       0.04 &       51.9/19 =       2.73 \\
$N_{200} = 4$ &        4.5 $\pm$        0.3 &       1.05 $\pm$       0.04 &       18.1/19 =      0.951 \\
$N_{200} = 5$ &        6.4 $\pm$        0.4 &       0.95 $\pm$       0.04 &       41.5/19 =       2.18 \\
$N_{200} = 6$ &        7.4 $\pm$        0.5 &       0.94 $\pm$       0.04 &       53.4/19 =       2.81 \\
$N_{200} = 7$ &        8.1 $\pm$        0.7 &       1.02 $\pm$       0.05 &       19.0/19 =       1.00 \\
$N_{200} = 8$ &       10.3 $\pm$        0.8 &       0.86 $\pm$       0.05 &       26.7/19 =       1.41 \\
$9 \le N_{200} \le 11$ &       12.5 $\pm$        0.6 &       0.93 $\pm$       0.03 &       26.0/19 =       1.37 \\
$12 \le N_{200} \le 17$ &       16.9 $\pm$        0.8 &       0.99 $\pm$       0.02 &       25.1/19 =       1.32 \\
$18 \le N_{200} \le 25$ &         23 $\pm$          1 &       0.99 $\pm$       0.03 &       38.9/19 =       2.05 \\
$26 \le N_{200} \le 40$ &         34 $\pm$          2 &       0.95 $\pm$       0.02 &       49.3/19 =       2.59 \\
$41 \le N_{200} \le 70$ &         47 $\pm$          2 &       0.89 $\pm$       0.03 &       48.9/19 =       2.57 \\
$71 \le N_{200} \le 220$ &         77 $\pm$          7 &       0.91 $\pm$       0.04 &       28.1/19 =       1.48
\enddata \tablecomments{Power law fits for each bin in \nvir: $\Delta\Sigma = AR^{-\alpha}$ with $R$ in units
of $h^{-1}$Mpc and $A$ in units of $h$M$_{\odot}$pc$^{-2}$.  The reduced $\chi^2$ is relatively high
in most bins, indicating a power law is not a good fit.  There is a strong correlation between
the power-law amplitude $A$ and richness \nvir, while the best fit $\alpha$ is relatively stable.
}
\end{deluxetable}

\begin{figure*}[h]
\centering
\plotone{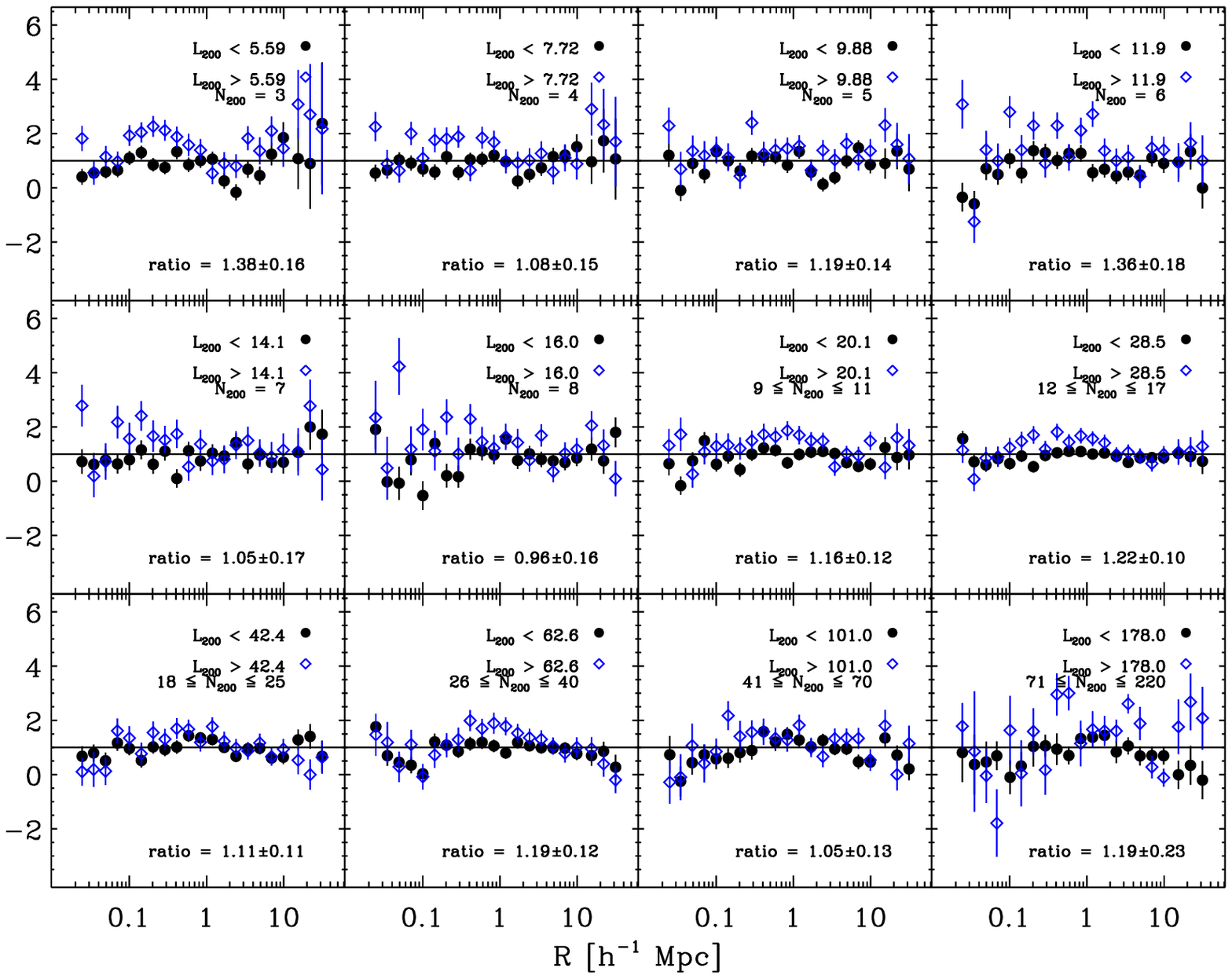}
\caption[\deltasig\ to 10 Mpc]{The same \numNgalBins\ \nvir\ bins from figure
  \ref{fig:deltasig_ngals200_12} with each bin split into two by \lvir.
  The \lvir\ split is at the 2/3 quantile.  The upper quantile is represented
  by the diamond symbols, the lower by the filled circles. The best fitting
  power-law for each \nvir\ bin before splitting is divided out. The mean ratio
  of the splits is shown in the legend. There is a further
  correlation with luminosity within each \nvir\ bin.
  \label{fig:deltasig_ngals200_ilum200}}
\end{figure*}

We split the \nvir\ sample further by \lvir\ within each bin as shown in Figure
\ref{fig:deltasig_ngals200_ilum200}.  In each of the \nvir\ bins we split at
the 2/3 quantile in luminosity, such that the 2/3 lowest objects are in one bin
and the top 1/3 are in the other. This quantile was chosen because any 
scaling with luminosity combined with the steep luminosity function would
predict equal S/N only for an uneven split.  In the figure, the best-fitting
power law has been divided out. In each case the upper \lvir\ quantile has a
stronger signal than the lower.  Although \nvir\ is correlated with the lensing
signal, there is additional information contained in \lvir.

\begin{figure*}[p] \centering
    \plotone{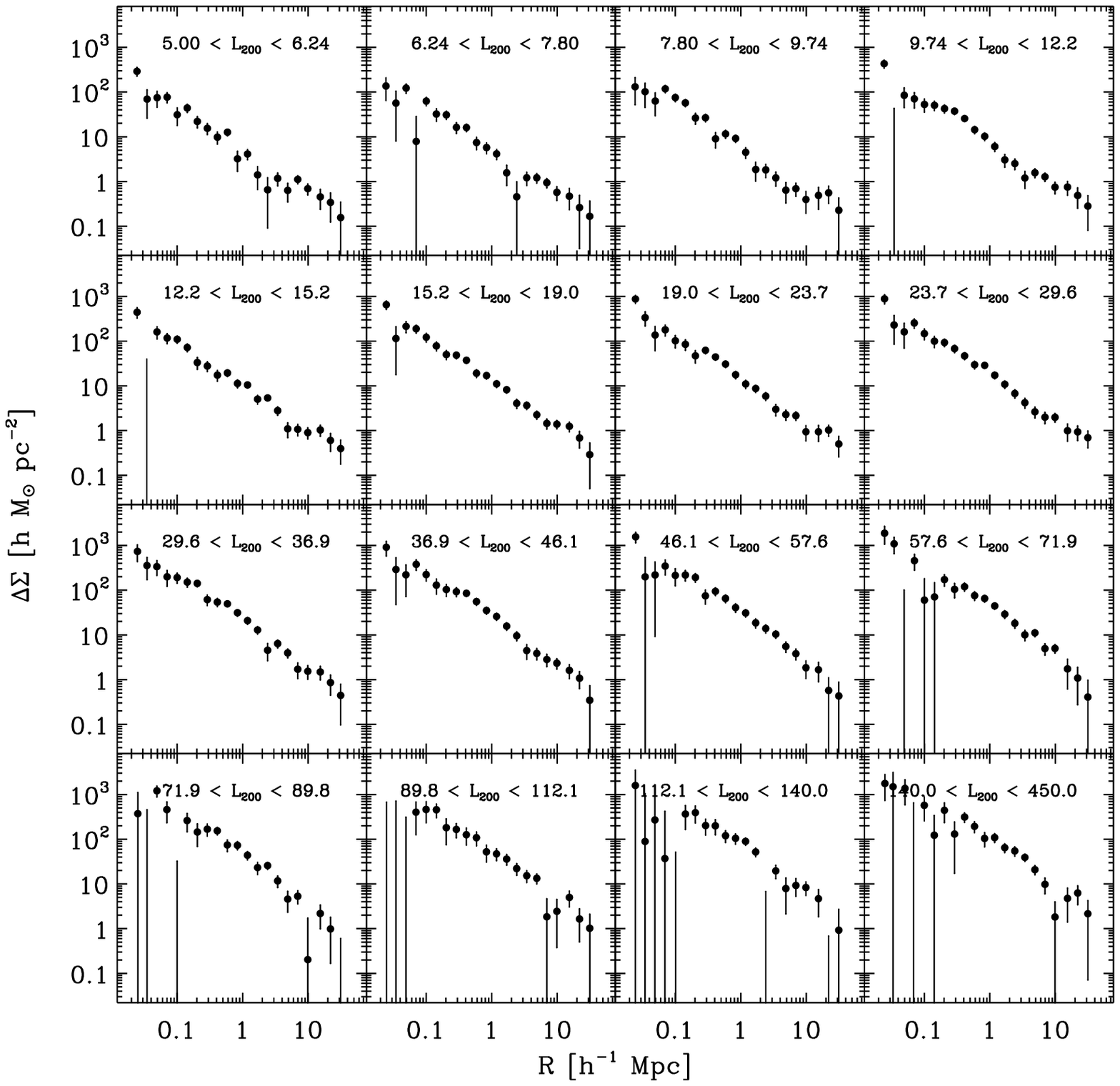}
    \caption[\deltasig\ in ilum200 bins]{Same as Figure
    \ref{fig:deltasig_ngals200_12} but in \numLumBins\ bins of \lvir, the total
    i-band luminosity of $(>0.4L_*)$ galaxies within \rvirgal.
    
    \label{fig:deltasig_ilum200_16}} \end{figure*}

We explored the luminosity dependence further by splitting the clusters
into \numLumBins\ bins by \lvir, without regard to the galaxy counts.  The
results of this binning are shown in Figure \ref{fig:deltasig_ilum200_16}.
We were able to split the sample into many more bins without losing significant
precision.  This is primarily due to splitting the lower \nvir\ bins into one
or more \lvir\ bins.  Generally the features are similar to the \nvir\ splits
but with more dynamic range in signal amplitude.  We also split each \lvir\ bin
into quantiles of \nvir, but saw no significant trend.

\begin{figure*}[p] \centering
    \plotone{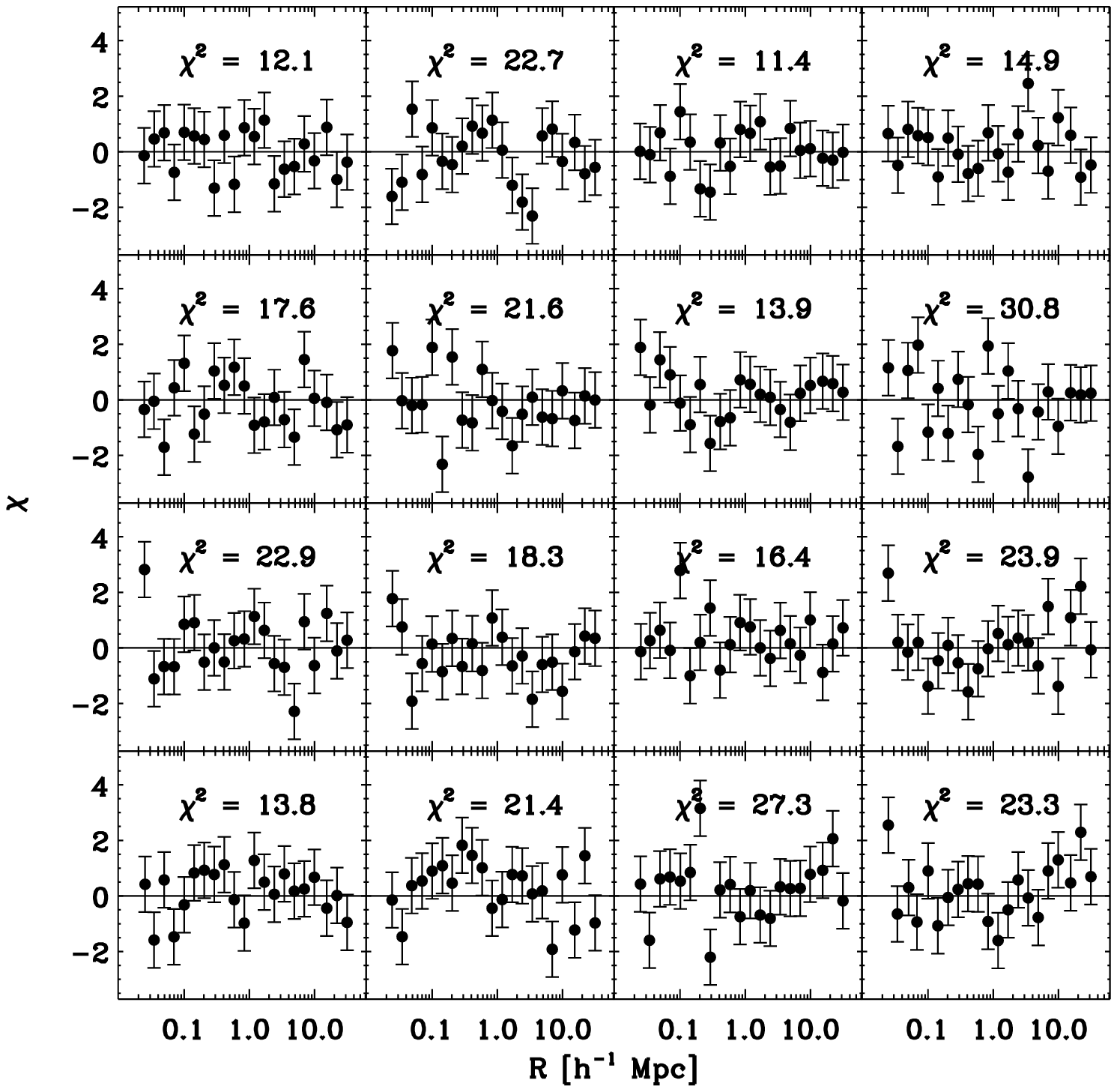} \caption[\deltasig\
    zregress]{Bins of luminosity as in Figure \ref{fig:deltasig_ilum200_16}
    with each \lvir\ bin split at the median redshift of 0.25.  The data points
    are the difference between high redshift and low redshift splits divided by
    the error (added quadratically). The distribution of $\chi^2$ between the
    \numLumBins\ bins is consistent with that expected from random scatter with
    21 degrees of freedom.  
    
    \label{fig:deltasig_ilum200_z}} \end{figure*}

Finally, we split the \lvir\ samples into bins of cluster redshift in order to
quantify any redshift evolution in the signal.  We split each bin at the median
redshift of 0.25.  The difference relative to the errors (combined
quadratically) is shown in Figure \ref{fig:deltasig_ilum200_z}.  No evolution
is evident.  The distribution of $\chi^2$ between the 16 \lvir\ bins is
consistent with that expected from random deviations for 21 degrees of freedom.
This test supercedes the weaker statement that the mean signal across all
clusters is consistent with no evolution.  Although not presented here for the
sake of brevity, we also have a null detection of redshift evolution in bins of
\nvir.

\epsscale{0.72}

\epsscale{1}

\section{Further Systematic Effects} \label{sec:systematics}

We have corrected the lensing measurements for additive errors and the
clustering of sources with lenses (see \S \ref{sec:corrections} ). In this
section we shall discuss other possible sources of systematic error.  Many of
these issues have been addressed in detail in \citet{MandelbaumSystematics05}.
We will briefly comment here on a few of the more important issues, in
particular the photometric redshifts for which our analysis differs from that
study, and intrinsic alignments in clusters.

\subsection{Lensing Measurement Errors} \label{sec:systematics:lensmeaserr}

There are two basic types of errors on lensing measurements: multiplicative
(calibration) errors and additive errors.  We described the corrections applied
for additive errors in \S \ref{sec:corrections}.

Calibration errors in \deltasig\ come in two types. Recall the
definition of \deltasig:
\begin{equation}
\Delta\Sigma(R) = \gamma_T (R) \times \Sigma_{crit}
\end{equation}
Errors may occur in converting the measured galaxy shapes to shear \gammat, or
in measuring the critical density \sigmacrit, which depends on the angular
diameter distances to lens and source.

Errors in the shear calibrations may be due to residuals in the PSF dilution
correction or incorrect shape to shear transformations.  The shape to shear
transformation is measured directly from the data (see \S \ref{sec:lensmeas}).
It is determined much better than our signal, so it should be a minimal
effect. The dilution corrections discussed in \S \ref{sec:data:shapes},
however, can only be determined in a model-dependent way. Thus it is difficult
to empirically determine the accuracy of the correction.  As a further test, we
split the source sample into bins of size and saw no variation of the recovered
the signal within the errors.  This however, does not demonstrate there is no
overall calibration error.

The calibrations were tested using simulations in the second Shear Testing
Programme, also known as STEP2 \citep{STEP2}.  For the PSFs tested therein, the
calibration was recovered to a few percent accuracy, but there is an inherent
uncertainty when any such scheme is applied to real data.  In particular, the
PSF in the STEP2 images are different from the typical SDSS PSF.  
%There is work
%in progress to test SDSS calibrations more directly \citep{MandelbaumTest}.

The PSF is not constant, but varies in time and position in the focal
plane. This variation is mapped by the measurement of a finite number of stars
per field.  The algorithm used by the SDSS to model the PSF
\citep{LuptonADASS01,Sheldon04}, while quite powerful, is inevitably limited in
the spatial frequency with which it can map changes in the PSF.  The density of
the stars used is about one per 6.7 square arcminutes, or a spacing of 2.6
arcminutes. Any PSF variations on scales smaller than this are simply not
accounted for. 

A recent study by \citet{JarvisJain04} has shown that this limitation can be
overcome if the PSF patterns are recurring and depend on only a few major
variables, such as position in the focal plane and focus.  We will explore this
method in future work.

The errors in estimating \sigmacrit\ come from errors in the estimating the
photometric redshifts.  We estimated these error distributions for our sample
and include them in our calculations (see \S \ref{sec:photoz}).   The essential
assumption in determining these biases is that the five band magnitudes of the
SDSS is sufficient information to determine the redshift distribution of a
given set of objects.  If this is not true, for example if surface brightness
or size matter and the training set differs from the photometric set in this
regard, the photozs may be biased.  This was tested in
\citet{MandelbaumSystematics05} by using a sample of luminous red galaxies for
which a well matched training sample can be used, and consistency was found at
the 10\% level.  We have not repeated that experiment for this data, but expect
the same level of consistency.  

On small scales the sources may be mis-measured due to the presence of the
brightest cluster galaxy, on which we center our lensing measurements.  In two
SDSS studies \citep{Masjedi06a,MandelbaumAstroPh06} it was shown that on
$\lesssim 20\arcsec$ scales the presence of bright luminous red galaxies (LRGs)
will bias the photometry and size measurement of neighboring galaxies
significantly.  This is essentially a calibration error because it may cause
errors in the photometric redshifts and dilution correction.  Because the BCGs
are similar to LRGs both in intrinsic properties and redshift distribution,
similar effects are expected to impact this study.  We have not, however,
repeated these tests for the current sample.  For this reason we do not correct
for such effects, but rather caution the reader that, for scales $\lesssim 50
h^{-1}$ kpc, the profiles presented in this work may be biased.  For
mass studies at the virial radius this is an insignificant volume and can be
safely ignored, but caution should be used for studies of the inner profile.

\subsection{Intrinsic Alignments} \label{sec:systematics:intrinsic}

We assume that the measured correlations between the shapes of the source
galaxies are due to gravitational lensing.  In practice the shapes of galaxies
themselves may be correlated.  In general this mostly washes out in a
measurement such as ours as long as the sources are far behind the lens. This
is because we correlate with the tangential frame around lenses and the sources
are at a large range of distances.  In practice, however, it is difficult to
identify a sample of sources which are all truly separate from the lenses, so
if intrinsic alignments exist the signal may indeed be contaminated.  Limits
from data \citep[see][and references therein]{HirataAlign04} indicate that this
is a small ($<15$\%) contaminant to the tangential shear for galaxy-galaxy
lensing surveys.  The contamination may be higher for clusters because a larger
number of source galaxies are actually physically associated with the lens.  

In order to test the effects of alignments, we used clusters determined from
SDSS spectroscopy as presented in \citep{Berlind06}.  We used the
volume-limited sample with absolute magnitude in the $r$-band less than -20.
There are 4119 clusters in this sample. For tracers of the intrinsic alignment
we use all galaxies from the SDSS ``main'' spectroscopic galaxy sample over the
same regions; there is no need for a volume-limited sample of shape tracers.
See \citet{Strauss02} for a description of the target selection algorithm.  The
use of only spectroscopic redshifts for clusters and tracers greatly reduces
the fraction of physically unassociated pairs. We chose all galaxies within our
search aperture and with velocities within $\pm 2000$ km/s.  No attempt was
made to match the luminosity distribution of the tracers to those of our actual
source galaxies. Shapes were corrected for PSF effects using the same
techniques described in \S \ref{sec:corrections}.

\begin{figure}[t]
%\epsscale{\epssmall}
\plotone{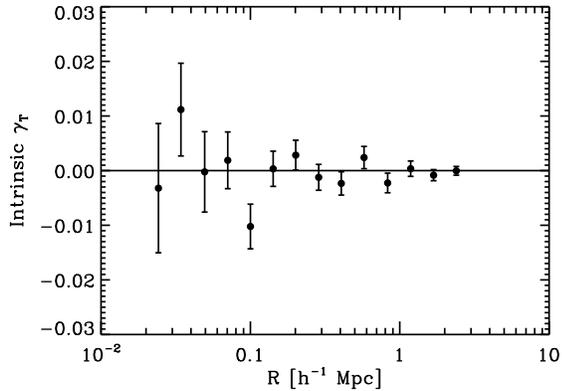}
  \caption{Estimate of intrinsic tangential alignments for the spectroscopic
    cluster sample of \citep{Berlind06}. The shear tracer population was drawn
    from the SDSS main spectroscopic sample and chosen to have velocity within
    $\pm 2000$ km/s of the cluster. Thus any signal is not due to lensing, but
    rather intrinsic correlations between galaxy shapes and the tangential
    direction relative to the lens center. Such an effect would bias the 
    lensing profiles.  No net effect was detected.
  \label{fig:intrinsic} }
%\epsscale{1}
\end{figure}

Figure \ref{fig:intrinsic} shows the mean tangential intrinsic shear measured
for these clusters.  There is no detected signal; we place only limits on the
effect.  For example, the intrinsic shear within 100 $h^{-1}$ kpc is $-0.0058 <
\langle \gamma_T \rangle < 0.0025$ at 95\% confidence.

\begin{figure*}[t]

    \epsscale{0.7} \centering
    \plotone{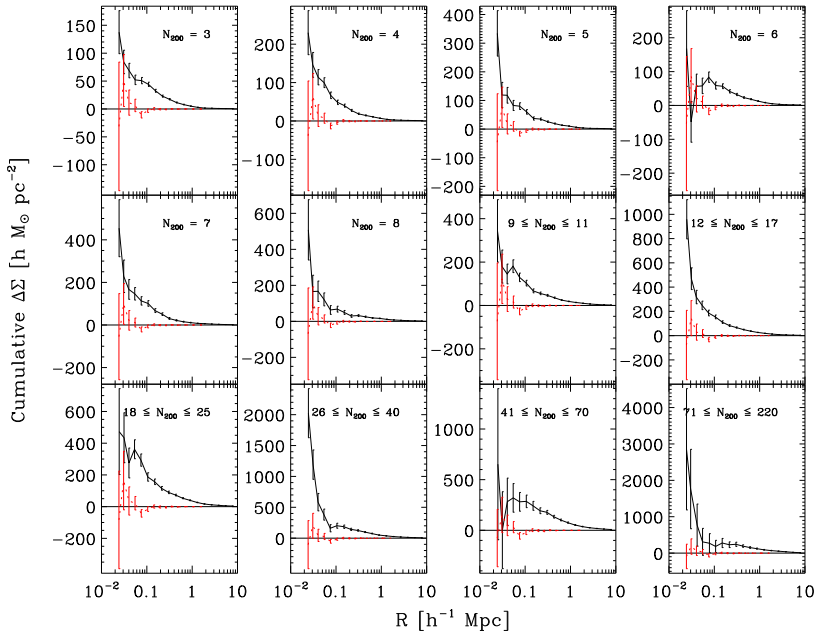}
    \caption{ Mean estimated contamination of \deltasig\ from intrinsic
    alignments in each of the \nvir\ bins, estimated as described in the text.
    These are cumulative curves so adjacent bins are correlated. In each
    panel, the intrinsic shear is the red curve and the cluster measurement is
    the black curve.     Note the vertical scale varies between
    plots.\label{fig:intrinsic_deltasig12}} 

\end{figure*}

In order to estimate the limit of contamination in our data we took the limits
of this shear and scaled them to \deltasig\ for each of our lens samples given
their mean redshift.  We also multiplied by the fraction of source galaxies
that were actually in the clusters, and the boost factor for clustering of
sources with the lenses.  This last factor is exactly that shown in
Figure \ref{fig:corr10Mpc}.  The mean estimated contamination is shown in
Figure \ref{fig:intrinsic_deltasig12} for each of the \nvir\ bins. This plot is
cumulative so adjacent points are correlated. Table \ref{tab:intrinsic} shows
limits on \deltasig\ in each of the \nvir\ bins within 100 $h^{-1}$ kpc.  These
limits are not very stringent but are all less than the mean signal in absolute
value.

\begin{deluxetable}{ccc}
\tabletypesize{\small}
\tablecaption{Intrinsic Alignment Limits for \nvir\ Bins \label{tab:intrinsic}}
\tablewidth{0pt}
\tablehead{
  \colhead{Bin}                                        &
  \colhead{$\Delta\Sigma_{int}$ Limits (95\%)}         &
  \colhead{$\langle \Delta\Sigma \rangle$} 
}
\
\startdata
$N_{200} = 3$ & [     -15.0,       4.03] &       39.6 $\pm$       4.05 \\
$N_{200} = 4$ & [     -20.1,       5.93] &       46.5 $\pm$       6.09 \\
$N_{200} = 5$ & [     -24.3,       6.12] &       56.8 $\pm$       8.30 \\
$N_{200} = 6$ & [     -27.9,       6.78] &       48.1 $\pm$       10.9 \\
$N_{200} = 7$ & [     -30.8,       8.26] &       73.0 $\pm$       13.6 \\
$N_{200} = 8$ & [     -34.3,       9.28] &       61.0 $\pm$       16.9 \\
$9 \le N_{200} \le 11$ & [     -39.8,       11.3] &       91.4 $\pm$       13.6 \\
$12 \le N_{200} \le 17$ & [     -49.4,       14.2] &       137. $\pm$       16.4 \\
$18 \le N_{200} \le 25$ & [     -65.5,       17.8] &       157. $\pm$       27.4 \\
$26 \le N_{200} \le 40$ & [     -73.5,       23.0] &       207. $\pm$       41.9 \\
$41 \le N_{200} \le 70$ & [     -91.7,       26.9] &       257. $\pm$       68.8 \\
$71 \le N_{200} \le 220$ & [     -120,       53.9] &       232 $\pm$       184 
\enddata 
\tablecomments{Limits on the contamination of \deltasig\ from
intrinsic alignments for each \nvir\ bin within a radius of 100 $h^{-1}$ kpc.
The limits are shown as 95\% confidence intervals.  For comparison, the
mean \deltasig\ for clusters within the same radius is listed in the last column.
}
\end{deluxetable}

The \citet{Berlind06} clusters are not selected in the same fashion as the
\maxbcg\ sample, so there is additional uncertainty in applying these limits.
Any sample selected to have spectroscopy will be selected differently, and will
be relatively few in number; this is a limit of the current observations.
Also, the effect may depend on richness, a possibility we are unable to address
due to the relatively small sample of \citet{Berlind06}.

As a further test of the effect of intrinsic alignments, we required the source
photoz in the \maxbcg\ shear measurements to be at least z(cluster) + 0.2, but
saw no change in the shear signal.  This could also mean that the redshifts are
dominated by random noise at the level of 0.2, but the mean error at these
magnitudes is expected to be $\lesssim 0.05$ (see \S \ref{sec:photoz})

\section{Complications for Interpretation} 

There are a few effects that may complicate the interpretation of these
results.  Firstly, the shear in the inner portions of the largest galaxy
clusters is not weak so interpretation of shear in terms of \deltasig\ is
incorrect. This is important for the largest clusters on $\lesssim 100 h^{-1}$
kpc scales.  This may be accounted for in a straightforward way when modeling
the signal \citep{MandelbaumAstroPh06,JohnstonLensing07}.  Secondly, the
cluster center, chosen as the location of the BCG, may not correspond to the
center of mass.  This acts like a convolution of the profile with the
distribution of BCG offsets.  Again, this is only important on relatively small
scales.  We will leave these issues for the follow-up paper
\citet{JohnstonLensing07}.  

A third issue is that of the richness measure \nvir.  There is a statistical
relationship between the number of galaxies counted by the \maxbcg\ cluster
finder \nvir\ and the true number of galaxies in the halo. This is partly due
to the fact that we observe the cluster in projection, so galaxies of the right
type and color but outside the cluster in the neighboring large scale structure
are counted.  It is also partly due to contamination of the counts due to
degeneracies in the space of galaxy type and redshift.  In this paper and in
the follow-up paper \citet{JohnstonLensing07} we choose, for simplicity, to
compare the lensing signal to the basic observable \nvir, and do not attempt to
account for the statistical relation between the measured counts and the true
counts.  In order to further interpret the measurements in terms of physical
models this relationship must be taken into account.  \citet{RozoCosmo07} model
this effect in their cosmological analysis of the \maxbcg.  We will perform a
similar analysis to that paper, this time including the lensing mass
calibrations, in \citet{RozoMassFunction07}.

\section{Summary}

We have measured ensemble lensing due to clusters of galaxies over scales 20
$h^{-1}$ kpc to 30 $h^{-1}$ Mpc. We split the sample into \numNgalBins\
independent bins of richness \nvir\ and \numLumBins\ bins of i-band luminosity
\lvir, with strong detections (S/N $\sim$15-20) in each bin.  The profiles were
corrected for systematic effects, including additive shear errors and
clustering of sources with the lens clusters.  We placed limits on the amount
of contamination in our signal due to intrinsic alignments and concluded that
on 100 kpc scales and greater the effect is not dominant; the limits on smaller
scales are weak. Calibration errors are less well known.  These calibration
errors are most likely dominated by uncertainties in determining the redshift
distribution of the sources, which are expected to be of order 10\%; the shear
calibration errors are expected to be a few percent. The shape of these
profiles, and relative scaling, are insensitive to calibration uncertainties.

Interpretation of these profiles on small scales requires some caution.  For
the most massive clusters, the lensing effect on $\lesssim 100 h^{-1}$ kpc
scales is most likely non-linear and this must be accounted for in models.
Furthermore, on $\lesssim 50 h^{-1}$ kpc (the first two bins) there may be
significant systematic effects in the photometry and size measurements of
source galaxies due to the extended light profile of the brightest cluster
galaxy. This effect has been seen around bright galaxies in the SDSS in other
studies \citep{Masjedi06a,MandelbaumAstroPh06}. Because the relative volume
interior to these radii is small, virial mass estimates should be robust, but a
study of the inner profiles will require further characterization of these
effects.  Neither of these issues should be important for scales $\gtrsim 100
h^{-1}$ kpc.  On the other hand, the BCG location chosen as the center may be
offset from the true mass peak and this can have effects to larger radius
\citep{JohnstonLensing07}.

The signal is dependent on richness and luminosity on all scales.  We fit power
law models and found the amplitude is a strong function of \nvir, while the
power law index is relatively insensitive to richness.  However, the signal is
a poor fit to a power law in most richness bins and the deviations from a power
law are systematic. The logarithmic slope generally runs from shallower to
steeper with increasing radius. We will interpret these curves in terms of a
more appropriate model with a universal halo profile and linear correlations on
large scales in \citet{JohnstonLensing07}.

Because the number of galaxies \nvir\ is not directly related to mass or
luminosity, one may expect a broad spread in luminosity and mass for a given
\nvir\ bin.  We explored this by splitting each \nvir\ bin into quantiles of
luminosity \lvir. We found a scaling of the signal with \lvir\ within
each \nvir\ bin, indicating that there is significant mass scatter in \nvir\
bins, and that mass may scale more strongly with \lvir\ than \nvir.

Finally, we explored the dependence of the signal on redshift by splitting each
of the luminosity bins at the mean redshift 0.25.  We detect no evolution
within our uncertainties for the current sample, though the redshift range
$0.1 < z < 0.3$ is relatively small. 

The precision of these measurements is sufficient to perform non-parametric
inversions to the mean three-dimensional mass density.  In the companion paper
\citet{JohnstonLensing07}, we present these inversions in each bin of richness
and luminosity. We then infer the model-independent virial mass and large scale
bias. We also interpret these profiles in terms of a universal halo profile on
small scales and linear bias on large scales.  In \citet{SheldonM2L07}, we
combine the non-parametric mass profiles from \citet{JohnstonLensing07} with
non-parametric light profiles to measure the mean mass-to-light ratios around
the \maxbcg\ clusters. In the forthcoming paper \citet{RozoMassFunction07}, we
use the mass-observable relation from \citet{JohnstonLensing07} to constrain
the mass function of halos and cosmological parameters.

\acknowledgements

    %\small

    Thanks to David Hogg and Michael Blanton for so many useful discussions,
    and to Roman Scoccimarro and Mulin Ding for use of the "Mafalda" 
    computing cluster.

    E.S. was supported by NSF grant AST-0428465.  B.K. and T.A.M. gratefully
    acknowledge support from NSF grant AST 044327 and the Michigan Center for
    Theoretical Physics

    The research described in this paper was performed in part at the Jet
    Propulsion Laboratory, California Institute of Technology, under a
    contract with the National Aeronautics and Space Administration.

    Funding for the SDSS and SDSS-II has been provided by the Alfred P. Sloan
    Foundation, the Participating Institutions, the National Science
    Foundation, the U.S. Department of Energy, the National Aeronautics and
    Space Administration, the Japanese Monbukagakusho, the Max Planck Society,
    and the Higher Education Funding Council for England. The SDSS Web Site is
    http://www.sdss.org/.

    The SDSS is managed by the Astrophysical Research Consortium for the
    Participating Institutions. The Participating Institutions are the American
    Museum of Natural History, Astrophysical Institute Potsdam, University of
    Basel, Cambridge University, Case Western Reserve University, University of
    Chicago, Drexel University, Fermilab, the Institute for Advanced Study, the
    Japan Participation Group, Johns Hopkins University, the Joint Institute
    for Nuclear Astrophysics, the Kavli Institute for Particle Astrophysics and
    Cosmology, the Korean Scientist Group, the Chinese Academy of Sciences
    (LAMOST), Los Alamos National Laboratory, the Max-Planck-Institute for
    Astronomy (MPIA), the Max-Planck-Institute for Astrophysics (MPA), New
    Mexico State University, Ohio State University, University of Pittsburgh,
    University of Portsmouth, Princeton University, the United States Naval
    Observatory, and the University of Washington.

    \normalsize

\newpage

\bibliographystyle{apj}
% Bib database
\bibliography{apj-jour,astroref}

\end{document}